\PassOptionsToPackage{usenames}{color}
\documentclass[12pt]{article}
\usepackage[draft=false]{hyperref}
\usepackage[normalem]{ulem}
\usepackage{lscape}
\usepackage{float}
\usepackage{xurl}
\usepackage{adjustbox}
\usepackage{multirow}
\usepackage{subcaption}
\usepackage{listings}
\usepackage{color}
\usepackage{xcolor}
\usepackage[bitstream-charter]{mathdesign}
\usepackage[a4paper, top=1in, bottom=1in, left=1in, right=1in]{geometry}

\usepackage{longtable}
\usepackage{graphicx}
\usepackage{soul}
\usepackage{tabularx}
\usepackage{tcolorbox}
\usepackage{bbm}
\usepackage{booktabs}
\usepackage{pgfplots}
\pgfplotsset{compat=1.3}
\usepackage{pdflscape}

\usepackage{array}
\usepackage{setspace}
\usepackage{dcolumn}
\usepackage{rotating}
\usepackage{amsmath, amsthm, mathtools}
\usepackage{enumitem}
\usepackage{verbatim}

\DeclareCaptionFont{mycap}{\fontsize{13}{15}\selectfont}
\usepackage[utf8]{inputenc}
\usepackage[T1]{fontenc}

\renewcommand*\thetable{\Roman{table}}

\usepackage[round]{natbib}
\usepackage{indentfirst} 
\usepackage[labelfont=bf,labelsep=period]{caption}   
\renewcommand{\tiny}{\fontsize{9}{9}\selectfont}

\hypersetup{colorlinks=true, allcolors=blue}

\pgfplotsset{ every non boxed x axis/.append style={x axis line style=-},
	every non boxed y axis/.append style={y axis line style=-},
	/pgfplots/ylabel near ticks/.style={
		/pgfplots/every axis y label/.style={
			at={(ticklabel cs:0.5)},rotate=90,anchor=near ticklabel}}
}

\pgfplotsset{
	tick label style={font=\small},
	xlabel style={font=\small},
	ylabel style={font=\small,align=center},
	legend style={font=\small,draw=none},
	title style={font=\small},
	legend cell align = left,
}

\theoremstyle{plain}

\theoremstyle{definition}

\newcommand{\PreserveBackslash}[1]{\let\temp=\\#1\let\\=\temp}
\newcolumntype{C}[1]{>{\PreserveBackslash\centering}p{#1}}

\renewcommand\arraystretch{1.5}
\usepackage{hyphenat}
\title{\LARGE Debiasing LLMs by Fine-tuning
}
\author{Zhenyu Gao, Wenxi Jiang, Yutong Yan\thanks{Gao, Jiang, and Yan are at the Department of Finance, CUHK Business School, The Chinese University of Hong Kong. Our correspondences are gaozhenyu@baf.cuhk.edu.hk, wenxijiang@baf.cuhk.edu.hk, and yutong.yan@link.cuhk.edu.hk, respectively. First draft: March 2026. A companion website is available at \url{https://debias-llm.yutongyan.xyz/}.
}
}
\date{April 2026}

\begin{document}

\maketitle
\linespread{1.5}
\begin{center}
  
\end{center}

{

\linespread{1.5}

\begin{abstract}
    \noindent Prior research shows that large language models (LLMs) exhibit systematic extrapolation bias when forming predictions from both experimental and real-world data, and that prompt-based approaches appear limited in alleviating this bias. We propose a supervised fine-tuning (SFT) approach that uses Low-Rank Adaptation (LoRA) to train off-the-shelf LLMs on instruction datasets constructed from rational benchmark forecasts. By intervening at the parameter level, SFT changes how LLMs map observed information into forecasts and thereby mitigates extrapolation bias. We evaluate the fine-tuned model in two settings: controlled forecasting experiments and cross-sectional stock return prediction. In both settings, fine-tuning corrects the extrapolative bias out-of-sample, establishing a low-cost and generalizable method for debiasing LLMs.
\footnotesize

\end{abstract}

}

\thispagestyle{empty}

\newpage
\linespread{1.5}

\setcounter{page}{1}
\renewcommand\arraystretch{1.5}

\newpage
\section{Introduction}

\noindent 
Algorithms and data-driven tools have already proven useful in financial decision-making~\citep[e.g.,][]{dAcunto2019promises, gu2020empirical}. More recently, AI agents built on large language models (LLMs) have been shown to simulate human behavior in economic experiments~\citep{horton2023large}.\footnote{Also see, e.g., \citet{brown2020language, bai2023qwen, chowdhery2023palm, openai2023gpt4, touvron2023llama} for foundational LLM architectures and \citet{yao2022react, park2023generative, wang2023voyager, schluntz2024agents} for LLM-based agents.} For these agents to be trusted with greater autonomy, however, the behavioral biases embedded in their underlying models become an economically important concern. 

A notable example is extrapolation bias, a well-documented tendency among human forecasts to place excessive weight on recent trends~\citep[e.g.,][]{da2021extrapolative, afrouzi2023overreaction}. 
Recent work shows that LLMs exhibit the same pattern: \citet{chen2024does} find that LLMs extrapolate trends when forecasting stock returns, and that this bias is difficult to correct. 
Prompt-based approaches, such as instructing the model to reason in a fully rational way, have little effect, suggesting the bias is encoded in the model's learned representations rather than driven by how the prompt is framed. 

To address this challenge, we develop a supervised fine-tuning (SFT) approach that could systematically mitigate behavioral biases in LLMs. 
A key insight is that if extrapolation bias arises from patterns learned during pretraining~\citep[e.g.,][]{gallegos2024bias}, then correcting it may require interventions beyond simple prompting.
To do so, we first construct instruction datasets: each prompt presents the model with a sequence of past data (e.g., stock returns), and the target response encodes a rational benchmark forecast. Then, we fine-tune on the instruction data using Low-Rank Adaptation~\citep[LoRA;][]{hu2022lora}, which directly reshapes how the model maps observed data to predictions. This replaces biased extrapolation patterns with rational forecasts while preserving the model's general language understanding.

Modern LLMs are built in two stages: (1) \textbf{pretraining} on a massive corpus of raw text to predict the next token in a sequence, and (2) \textbf{alignment} where the model is fine-tuned on curated examples and human feedback to produce conversational assistants. The pretraining corpus is rich in financial content: journalism, analyst reports, earnings call transcripts, and investment forums, where extrapolative language about firm performance is pervasive. Because the model learns by internalizing the statistical patterns in this text, its parameters encode not only factual knowledge, but also the systematic biases present in the data. Alignment shapes how the model communicates, including its tone, safety, and instruction-following, but does not correct its underlying beliefs about how data are generated. As a result, prompting-based approaches fail to mitigate LLM forecasting biases as they operate entirely at inference time without touching the model's parameters.  

Our framework introduces an additional fine-tuning step after alignment, directly targeting the model's forecasting behavior before deployment. We maintain a strict separation between training, validation, and test data. To establish a baseline, we first present the model with a held-out set of forecasting prompts and record its raw predictions. Each prompt provides a series of data points or a history of past stock returns and asks the model to forecast the next period, with no guidance on how to form that forecast. The model simply produces whatever prediction its pretrained weights generate. Comparing these predictions against rational benchmarks reveals the systematic biases we aim to correct. These prompts are reserved as the test set and never exposed to the model during training.

We then construct a separate instruction dataset of prompt-response pairs. The prompts have the same structure as those in the test set but use disjoint realizations, while the responses encode rational forecasts derived either from a rational expectations benchmark or from realized future returns. Each pair is a corrective example: it presents the model with the same input it already sees, but pairs it with the answer a disciplined forecaster would give. A portion of this instruction dataset is also held out as a validation set, used to monitor generalization during training and to determine the stopping rule.

A key technical choice is how to update the model's parameters. In earlier, smaller models such as BERT-large (340 million parameters), the standard approach was full fine-tuning: updating every weight on task-specific data~\citep{devlin2019bert, huang2023finbert}. 
Modern LLMs, however, are orders of magnitude larger. Qwen3-32B, the model we use, contains 32 billion parameters, making full fine-tuning challenging for two reasons. 
First, it is computationally prohibitive, requiring hundreds of gigabytes of GPU memory. 
Second, and more importantly, it risks catastrophic forgetting~\citep{goodfellow2013empirical}: updating all model parameters can lead to degradation in general capabilities. Since our goal is to correct a specific forecasting bias while preserving all other capabilities, full fine-tuning is poorly suited to the task.

We instead fine-tune using Low-Rank Adaptation~\citep[LoRA;][]{hu2022lora}, the most widely used parameter-efficient fine-tuning method in machine learning. Rather than updating the full model, LoRA freezes the original pretrained weights and attaches a small set of trainable parameters to each layer. Only these added parameters are updated during training, and because they typically represent less than 1\% of the full model, LoRA reduces computational cost by several orders of magnitude. More importantly, keeping the original weights frozen preserves the model's general language understanding while adjusting only the forecasting behavior of interest. Once training is complete, the added parameters are merged back into the original weights, so the fine-tuned model can be deployed with no additional inference overhead. We use early stopping based on validation performance to guard against overfitting.

We implement the framework on Qwen3-32B~\citep{yang2025qwen3}, an open-weight LLM with 32 billion parameters. We choose an open-weight model deliberately: unlike proprietary LLMs accessed via APIs, open-weight models allow researchers to inspect and modify internal parameters, which is a prerequisite for our fine-tuning approach. The procedure is also computationally inexpensive. The entire training costs a few hundred dollars on a commercial cloud cluster, negligible relative to the millions of dollars required to pretrain a frontier LLM from scratch.

We validate our fine-tuned model in two settings. The first replicates the controlled forecasting experiment of \citet{afrouzi2023overreaction}, replacing human subjects with the LLM. In that study, participants are assigned to one of six AR(1) processes with persistence $\rho \in \{0.0, 0.2, 0.4, 0.6, 0.8, 1.0\}$, observe 40 historical realizations, and submit one- and two-period-ahead forecasts over 40 rounds. Overreaction is quantified by regressing forecast errors on forecast revisions: a negative coefficient ($\hat{b}$) indicates that upward revisions systematically predict negative errors, a sign of overreaction.

We adapt this design for a text-based LLM. The off-the-shelf Qwen model closely replicates the pattern found in the human data: the overreaction coefficient $b$ is negative and statistically significant at the 1\% level across all six persistence conditions, most negative at $\rho = 0.0$ ($\hat{b} = -0.456$, $t = -19.05$) and monotonically less so as persistence increases, reaching $\hat{b} = -0.260$ ($t = -10.37$) at $\rho = 1.0$. Overreaction is strongest for transitory processes and weakest for random walks, exactly as observed in the human data.

We then fine-tune on a training dataset of approximately 30,000 round-level observations, constructed from 128 independent AR(1) realizations for each of the six persistence values, using conditional expectations as learning targets rather than ex-post realizations. After fine-tuning, the overreaction bias becomes statistically insignificant: point estimates of $b$ range from $-0.073$ ($t = -1.54$) at $\rho = 0.0$ to $-0.027$ ($t = -0.97$) at $\rho = 1.0$.

The second exercise turns to forecasting actual stock returns. Prior studies document a pronounced extrapolation bias in this domain: human forecasts~\citep{da2021extrapolative} and \mbox{ChatGPT-4}~\citep{chen2024does} alike place excessive weight on recent past returns. We adopt a similar design and prompt the model to forecast monthly returns for S\&P 500 constituents using trailing twelve-month return histories. We anonymize the prompt, supplying only numerical return histories with no firm or date identifiers, to mitigate lookahead bias. We quantify extrapolation by regressing the model's forecasts on lagged stock returns. The baseline model loads positively on all lags, with the coefficient on the most recent-month return equal to $0.394$ ($t = 53.92$) and declining with lag length, confirming excessive weight on recent performance.

To implement the SFT procedure, we divide the sample into three non-overlapping periods following \citet{gu2020empirical}: training (January 2001--December 2011), validation (January 2012--December 2015), and test (January 2016--December 2024). Unlike the previous exercise, we use realized next-month returns as the learning target, since the conditional expected return is not directly observable. By training on the empirical distribution of monthly returns conditional on trailing histories, the model learns the short-term reversal pattern in stock returns. After fine-tuning, the extrapolative loading is corrected: coefficients on all lagged returns turn negative, with the most recent lag equal to $-0.120$ ($t = -23.21$). The fine-tuned model has internalized from the training data that recent outperformers tend to reverse, producing forecasts that better reflect the actual return-generating process.

The correction holds out-of-sample in both the controlled experimental and empirical settings, ruling out the possibility that the behavioral change is an artifact of in-sample fitting. More broadly, our approach offers a low-cost and generalizable method for aligning LLM behavior with rational benchmarks across economic settings. This is a prerequisite for the responsible deployment of AI agents in financial decision-making. 

Our results carry direct implications for the next generation of automated financial advice. 
First-generation robo-advisors, which rely on rules-based portfolio optimization, have already been shown to reduce pervasive behavioral biases such as the disposition effect and trend chasing among retail investors~\citep{dAcunto2019promises}. 
As robo-advisory platforms increasingly integrate LLM-based agents to generate return forecasts from news and analyze earnings reports~\citep[e.g.,][]{lopez2023can, jha2024chatgpt}, the biases embedded in these models can harm advice quality. For example, an LLM agent that extrapolates recent trends into its forecasts would undermine the very discipline that makes automated advice attractive: clients would receive recommendations that amplify, rather than attenuate, the behavioral biases they sought to avoid. Our fine-tuning approach offers a practical remedy, enabling platform developers to debias the forecasting layer of an LLM-powered advisor before deployment. 

Beyond robo-advising, debiased LLM forecasts are relevant wherever autonomous agents act on predictions derived from historical patterns. In credit risk assessment, an extrapolative model may overweight a borrower's recent repayment trajectory and underestimate mean reversion in default risk, leading to procyclical lending decisions~\citep[e.g.,][]{feng2023empowering}. In macroeconomic nowcasting, extrapolation bias can amplify transitory shocks, producing misleading signals for policymakers~\citep[e.g.,][]{li-etal-2024-econagent}. In algorithmic trading and prediction markets, it can cause LLM-based signals to chase recent price trends rather than anticipate reversals~\citep[e.g.,][]{kim2026llm}. In each of these settings, the bias operates through the same channel we document: the model's pretrained parameters encode a tendency to extend recent patterns forward, and correcting this tendency requires intervention at the parameter level. Our SFT framework is directly applicable to these domains, as it requires only that the practitioner can specify a rational benchmark or a realized outcome against which to train.

\subsection*{Related Literature}
\paragraph{Use of LLMs in Finance}
A growing body of work applies LLMs to extract economically meaningful signals from financial text. Studies in this vein use corporate disclosures and earnings calls to predict firm actions~\citep[e.g.,][]{cao2023talk, jha2024chatgpt}, financial news to forecast stock returns~\citep[e.g.,][]{chen2022expected, lopez2023can}, central bank communications to classify policy stance and identify macroeconomic shocks~\citep[e.g.,][]{hansen2024can}, historical news to generate economic expectations~\citep{bybee2023ghost}, mutual fund manager reports and analyst research to study belief formation~\citep[e.g.,][]{gao2024structured, ke2024analysts}, and published books to measure popular sentiment toward finance~\citep{jha2025does}. More broadly, recent work shows that GPT can serve as an accurate measurement tool for qualitative attributes across a wide range of domains, performing on par with human evaluators~\citep{asirvatham2026gpt}. %
Beyond using LLMs as text processors, a parallel strand deploys them as simulated or autonomous agents capable of replicating experimental economics findings, generating emergent market dynamics in simulated trading environments, nowcasting stock returns from autonomously gathered information, and automating the discovery of novel economic theories~\citep[e.g.,][]{horton2023large, lopez2025can, chen2026autonomous, lopez2026can}. 
A separate line of work seeks to open the LLM black box: \citet{chen2024out} show that next-token conditional probabilities provide an interpretable measure of prediction uncertainty that improves portfolio performance, while \citet{chen2025financial} apply sparse autoencoders~\citep{lieberum2024gemma} to extract interpretable concepts from LLM embeddings and steer outputs along interpretable feature dimensions such as risk aversion and optimism.

Our work instead intervenes after alignment, using SFT to directly modify how a Chat-LLM forms its beliefs.

\paragraph{LLM biases in behavioral economics.}
Most closely related to our work is the emerging literature examining whether LLMs inherit human-like behavioral biases.
\citet{chen2024does} show that LLMs over-extrapolate when forecasting stock returns and that prompt-based interventions, such as role-based instructions, have a negligible effect on this bias. 
\citet{bini2025behavioral} conduct a comprehensive battery of behavioral economics experiments and find that while rational prompting can modestly reduce biases in some settings, it does not fully eliminate them. Our paper focuses on the stock return forecasting setting of \citet{chen2024does}, where prompting is least effective, and shows that parameter-level intervention through SFT can succeed where prompting fails. Other studies suggest that LLMs display partially human-like economic behavior and can exhibit behavioral biases under certain conditions~\citep{fedyk2024ai, ross2024llm, wu2025llm}.

Our SFT approach contributes to this literature by showing that LLM biases can be effectively corrected through targeted parameter-level intervention. The method is low-cost and potentially generalizable to aligning LLM behavior with rational benchmarks across a broad range of economic settings.

\section{Methodology}
\subsection{Fine-Tuning LLMs}

Figure~\ref{fig:pretraining_pipeline} summarizes the modern LLM training pipeline. To fix ideas, consider the progression from GPT-3~\citep{brown2020language} to ChatGPT-3.5. The pipeline has two phases: a training phase, in which the model learns from data, and a prompting phase, in which users interact with the trained model.

\paragraph{Pretraining.} Training begins with a massive corpus of raw text, typically tens of terabytes drawn from web crawls, books, academic papers, and other sources. Critically for our purposes, this corpus includes financial journalism, analyst commentary, earnings call transcripts, and online investment forums, all sources in which extrapolative language about asset returns is pervasive. A neural network~\citep{vaswani2017attention} is trained on this corpus via next-token prediction: given a sequence of words, the model learns to predict the next word. This objective is entirely self-supervised, requiring no human labels or curated examples. By predicting the next token across trillions of training examples, the model internalizes the statistical regularities of its training data, including factual knowledge, reasoning patterns, and, unavoidably, any systematic biases present in the text. For example, GPT-3 was trained on approximately 570~GB of filtered text, producing a \textit{base model} capable of generating fluent text but not designed to follow instructions or engage in dialogue.

\paragraph{Alignment.} The base model is then fine-tuned to produce helpful, instruction-following responses. This alignment stage uses curated prompt-response pairs and human preference data~\citep{ouyang2022training} to teach the model the format and style of a conversational assistant.\footnote{Common preference optimization methods include direct preference optimization~\citep[DPO;][]{rafailov2023direct} and group relative policy optimization~\citep[GRPO;][]{shao2024deepseekmath}.} The aligned model, often referred to as a Chat-LLM, is the version deployed to end users (e.g., ChatGPT-3.5). This two-stage process, pretraining on raw text followed by alignment to human preferences, is now the standard paradigm across all major LLM families. Importantly, alignment shapes \textit{how} the model responds, including its tone, safety, and adherence to instructions, but does not target the \textit{substance} of its beliefs about data-generating processes or economic relationships. Moreover, because the human feedback used during alignment reflects the judgments of human annotators, who themselves may hold extrapolative beliefs about asset returns, the alignment stage can reinforce or even amplify biases already present in the base model. Extrapolation bias in the deployed Chat-LLM may therefore originate from both stages of training: absorbed from financial text during pretraining and reinforced through human preferences during alignment.

\paragraph{Prompting.} At inference time, users interact with the aligned Chat-LLM through natural language prompts. This is the stage at which all existing attempts to mitigate LLM forecasting biases operate: by modifying the prompt (e.g., role-based instructions, few-shot demonstrations, chain-of-thought reasoning) without altering the model's parameters. As documented by \citet{chen2024does}, prompt-level interventions have limited efficacy against extrapolation bias. Because this bias is encoded in the model's parameters during both pretraining and alignment, it cannot be corrected by reframing the input alone.

\paragraph{Our approach.} The central insight is that correcting a bias embedded in the model's parameters requires intervening at the parameter level. Since researchers and practitioners typically have access only to the final Chat-LLM, not the pretraining corpus or the alignment data, the only feasible point of intervention is an additional fine-tuning step applied to the Chat-LLM itself. We introduce such a step between the standard alignment phase and deployment (Figure~\ref{fig:sft_pipeline}). Starting from the aligned Chat-LLM, we apply supervised fine-tuning (SFT) on a curated dataset of prompt-response pairs in which each prompt presents a return history and each response contains the corresponding rational benchmark forecast. SFT teaches the model to produce rational forecasts in place of its default over-extrapolative predictions, directly reshaping the mapping from observed data to predictions. Section~\ref{sec:debiasing_framework} describes the full framework.

\subsection{Debiasing Framework}\label{sec:debiasing_framework}

Our debiasing framework rests on a simple intuition: if an LLM systematically overreacts to recent returns or extrapolates transitory trends, we can teach it not to. The key is to show the model what rational forecasts look like and let it learn the correction itself. We implement this idea through supervised fine-tuning (SFT), maintaining a strict separation between the data used to train the model and the data used to evaluate whether debiasing has succeeded. Figure~\ref{fig:debiasing_framework} summarizes the framework.

\paragraph{Bias identification.} We begin by identifying the bias. We present the baseline LLM with a held-out set of forecasting prompts and collect its raw predictions. Each prompt supplies the model with a history of past returns and asks it to forecast the next one or two periods ahead. The model receives no guidance on how to form its forecast; it simply produces whatever prediction its pretrained weights generate. Comparing these predictions against rational benchmarks, whether derived from a rational expectations model or from ex-post realizations, exposes the systematic biases we aim to correct. A model exhibiting overreaction, for instance, will chase recent momentum too aggressively: after a sequence of negative returns, it predicts further steep declines even when the data-generating process implies mean reversion. A model exhibiting extrapolation will latch onto short-run patterns and project them forward as if they were permanent, ignoring the transitory nature of the underlying shocks. This diagnostic phase pins down both the direction and the severity of the distortion. Critically, the prompts used in this phase are set aside as our test set and are never exposed to the model during training.

\paragraph{Instructional dataset.} The central ingredient of the framework is the instructional dataset. We construct a separate collection of prompt-response pairs in which the prompts share the same structure as those in the test set, presenting the model with a return history and asking for a forecast, but the responses now reflect what a rational forecaster would predict. These target responses can be sourced in two ways: from the conditional forecasts implied by a rational expectations benchmark, or from the realized future returns that the model is trying to predict. In either case, the target encodes the behavior we want the model to internalize. Where the baseline LLM might dramatically overweight a recent crash and forecast continued freefall, the instructional response instead shows a measured, mean-reverting prediction. Where the baseline might project a brief rally into an extended boom, the instructional response shows a tempered, stabilizing path. In essence, each prompt-response pair is a corrective example, pairing the same information the model already sees with the answer a disciplined forecaster would give. We further partition a portion of this instructional data into a validation set, which plays no role in updating model weights but is used to track generalization performance during training and to inform our stopping rule. Together, the training and validation splits constitute the data the model learns from, entirely disjoint from the test set reserved for final evaluation.

\begin{figure}[H]
\centering
\caption{LoRA Reparametrization}
\label{fig:lora_paper_figure1}
\vspace{0.2cm}
{\fontsize{11}{15.6}\selectfont
\parbox{\textwidth}{
This figure from \citet{hu2022lora} illustrates the LoRA reparametrization. The pretrained weights (left) are frozen during fine-tuning. A small pair of trainable matrices (right, in orange) is attached alongside the frozen weights to capture the task-specific adjustment. Only these added parameters are updated during training, typically less than 1\% of the full model, making fine-tuning feasible.
}
}
\bigskip
\includegraphics[width=0.6\textwidth]{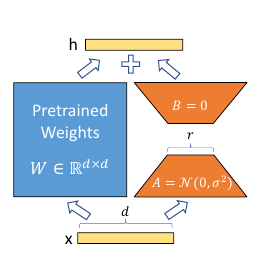}
\end{figure}

\paragraph{Fine-tuning with LoRA.} With the instructional dataset in hand, we fine-tune the LLM using Low-Rank Adaptation~\citep[LoRA;][]{hu2022lora}. Rather than updating all of the model's parameters, LoRA freezes the original pretrained weight matrices and introduces a parallel low-rank update (Figure~\ref{fig:lora_paper_figure1}). Specifically, for each pretrained weight matrix $W_0 \in \mathbb{R}^{d \times k}$ in the model's attention layers, LoRA adds two smaller matrices: a down-projection $A \in \mathbb{R}^{r \times k}$ and an up-projection $B \in \mathbb{R}^{d \times r}$, where the rank $r \ll \min(d, k)$. For a given input $x \in \mathbb{R}^k$, the layer output is computed as $h = W_0 x + BAx$. Here $x$ is a numerical vector representing a single token at a particular layer of the network. As the model processes a prompt, it first splits the text into tokens (subword units, which may be words, parts of words, or numbers) and converts each token into such a vector. The weight matrix $W_0$ transforms each token's representation to produce the layer's output. LoRA supplements this transformation with the low-rank term $BAx$, so the model can adjust its behavior without altering $W_0$ itself. Because $B$ is initialized to zero, the product $BA$ is zero at the start of training, so the model begins with exactly the same behavior as the pretrained model. As training progresses, only $A$ and $B$ are updated, allowing the model to learn a task-specific adjustment $\Delta W = BA$ without modifying the original weights. At inference time, the learned matrices can be merged directly into the pretrained weights by computing $W' = W_0 + BA$, introducing no additional latency compared to a standard model. This design choice serves two purposes. First, it is computationally efficient: fine-tuning billions of parameters from scratch would be prohibitively expensive, whereas LoRA requires only a fraction of the memory and compute\footnote{Effective training size is usually less than 1\% of the full model.}. Second, and more important for our setting, it preserves the LLM's general language understanding while surgically adjusting only the forecasting behavior of interest. The model retains its capacity to parse numerical inputs, follow instructions, and produce coherent outputs; what changes is the mapping from observed return histories to predicted future returns. Throughout training, we monitor the loss on the training set alongside the model's forecasting performance on the validation set. We employ early stopping: once validation performance ceases to improve, we halt training and select the checkpoint with the best validation performance. This guards against overfitting to the idiosyncrasies of the training sample and ensures that the debiasing generalizes beyond the specific examples the model has seen.

\paragraph{Out-of-sample evaluation.} The final and most important test is whether the debiased model generalizes to data it has never encountered. We return to the held-out test set from the diagnostic phase and feed the same prompts to the fine-tuned LLM. Because these prompts were excluded before training began, the model's responses to them constitute a clean, out-of-sample evaluation. We then compare the debiased model's forecasts against the same rational benchmarks used in the diagnostic phase. If fine-tuning has worked, the gap between the model's predictions and the rational forecasts should narrow substantially: overreaction to recent returns should be attenuated, extrapolation of transitory patterns should diminish, and the overall distribution of forecast errors should shift toward the behavior implied by the benchmarks. This out-of-sample evaluation discipline is what allows us to claim that the debiasing is genuine, a learned change in how the model processes return histories, rather than an artifact of in-sample fitting to the particular sequences it trained on.

\subsection{Implementation Details}

\paragraph{Model choice.} We implement the debiasing framework on Qwen3-32B~\citep{yang2025qwen3}, an open-weight LLM with 32B parameters, as it is the most popular dense model with over 30B parameters and offers the best balance between capability and accessibility among the open-source models. We choose an open-weight model deliberately: unlike proprietary LLMs accessed through APIs, open-weight models allow researchers to inspect, modify, and retrain the model's internal parameters, which is a prerequisite for our fine-tuning approach. Qwen3-32B offers a strong balance between forecasting capability and tractability; it is large enough to exhibit the sophisticated behavioral patterns we aim to correct, yet small enough to fine-tune efficiently with LoRA.

\paragraph{Training procedure.} After downloading the pretrained model weights, we attach LoRA adapter layers to the model's attention modules. These adapters introduce a small number of trainable parameters while leaving most of the parameters entirely frozen. The model then undergoes supervised fine-tuning on our instructional dataset: at each training step, the model receives a forecasting prompt, generates a prediction, and updates only the adapter weights to bring that prediction closer to the rational target response. We monitor performance on the validation set throughout training and apply early stopping to select the checkpoint at which debiasing is strongest without sacrificing out-of-sample generalization.

\paragraph{Software and compute.} We implement the training pipeline using standard open-source libraries for model loading, tokenization, training orchestration, and LoRA integration. The computational cost is negligible relative to the millions of dollars typically required to pretrain a frontier LLM from scratch. 

\paragraph{Inference.} For prompting, we use \texttt{vLLM}~\citep{kwon2023efficient}, an open-source, high-throughput inference library designed to efficiently process large volumes of prompts in parallel. Because our experimental design requires eliciting forecasts across a large number of prompts, often numbering in the tens of thousands, efficient inference is essential to keeping the overall pipeline tractable. We set the sampling temperature to zero to produce deterministic outputs.\footnote{Greedy decoding with zero temperature selects the highest-probability token at each step, but floating-point operations on GPUs are not perfectly associative. Changes in batch size alter the order of parallel computations within the attention mechanism, introducing small numerical discrepancies that can occasionally shift which token ranks highest. This is a well-documented property of parallelized GPU inference rather than a limitation specific to our setting. We verify that our qualitative results are robust to batch-size variation.}

\section{Results}
\subsection{Controlled Experiments}
\paragraph{Experimental design.} In this exercise, we generally follow the approach of~\cite{afrouzi2023overreaction}, with the difference that we use an LLM as our participant, while they recruit humans. \citet{afrouzi2023overreaction} design a controlled forecasting experiment to measure biases in expectations.

In their baseline experiment, they recruit approximately 207 participants from Amazon's Mechanical Turk platform and randomly assign each to one of six AR(1) processes,
\begin{equation}\label{eq:ar1}
    x_t = \rho \, x_{t-1} + \varepsilon_t, \qquad \varepsilon_t \overset{i.i.d.}{\sim} \mathcal{N}(0, \sigma^2),
\end{equation}
with persistence $\rho \in \{0.0, 0.2, 0.4, 0.6, 0.8, 1.0\}$, a long-run mean of zero, and a conditional volatility of $\sigma = 20$.
Each participant first observes 40 historical realizations of the assigned process displayed as a time series graph.
The forecasting game then proceeds for 40 rounds.

At each round~$t$, participant~$i$ sees the realized history $x_{t-39}, \ldots, x_{t-1}, x_t$ and submits two forecasts: a one period ahead forecast $F_{i,t}\, x_{t+1}$ and a two period ahead forecast $F_{i,t}\, x_{t+2}$.
The true value $x_{t+1}$ is then revealed, the graph is updated, and round~$t+1$ begins.
Because participants forecast both one and two periods ahead in every round, two forecasts of the same target~$x_{t+1}$ are collected from each participant across consecutive rounds: the two period ahead forecast $F_{i,t-1}\, x_{t+1}$ submitted in round~$t-1$ (when $x_{t-1}$ was the latest observation), and the one period ahead forecast $F_{i,t}\, x_{t+1}$ submitted in round~$t$ (after $x_t$ is observed).
The difference $F_{i,t}\, x_{t+1} - F_{i,t-1}\, x_{t+1}$ is the forecast revision, capturing how participant~$i$ updated the prediction of~$x_{t+1}$ upon seeing the new realization~$x_t$.
The forecast error $x_{t+1} - F_{i,t}\, x_{t+1}$ is realized once $x_{t+1}$ is revealed at the start of round~$t+1$.

To quantify overreaction, the authors estimate, for each level of~$\rho$, the following panel regression:
\begin{equation}\label{eq:err_rev}
  \underbrace{x_{t+1} - F_{i,t}\, x_{t+1}}_{\text{forecast error}}
  \;=\; a \;+\; b\,\underbrace{\bigl(F_{i,t}\, x_{t+1} - F_{i,t-1}\, x_{t+1}\bigr)}_{\text{forecast revision}}
  \;+\; v_{i,t},
\end{equation}
A negative coefficient~$b$ indicates that upward revisions systematically predict negative forecast errors, which is the signature of overreaction.

\paragraph{LLM replication.} We take this experimental framework as our starting point and replicate it with Qwen3-32B serving as participants.
Since the original experiment relies on a visual interface in which participants observe and interact with a time series graph, we adapt the design for text-based language models that do not process visual input.
Specifically, at each round~$t$, we provide the model with the numerical values of the past realizations $x_{t-39}, \ldots, x_{t-1}, x_t$ and ask it to forecast the changes in $x_{t+1}$ and $x_{t+2}$.
A notable feature of our prompt design is that we elicit predicted \emph{changes} rather than predicted \emph{levels}. 
This choice is motivated by the cognitive process that the original experiment documents: when human subjects observe a time series graph, they anchor on the last realized value and form a judgment about the magnitude and direction of the next movement. 
Eliciting changes from the LLM parallels this anchoring process, prompting the model to condition explicitly on the most recent realization before producing each forecast. 
We otherwise retain the same process parameters as \citet{afrouzi2023overreaction}, setting the long-run mean to zero, the conditional volatility to~20, and the persistence to each of $\rho \in \{0.0, 0.2, 0.4, 0.6, 0.8, 1.0\}$.
For each value of~$\rho$, we sample a set of independent AR(1) realizations as our test set.

\begin{figure}[H]
\centering
\caption{Controlled Experiment: Base Model and Prompt-Based Debiasing}
\label{fig:qje_table1_replication}
\vspace{0.2cm}
{\fontsize{10}{13}\selectfont
\parbox{\textwidth}{
This figure plots the estimated overreaction coefficient $b$ from the error-revision regression $x_{t+1} - F_{i,t}\, x_{t+1} = a + b\,(F_{i,t}\, x_{t+1} - F_{i,t-1}\, x_{t+1}) + v_{i,t}$ for each persistence parameter $\rho \in \{0.0, 0.2, 0.4, 0.6, 0.8, 1.0\}$, using forecasts from Qwen3-32B. Panel~A uses the baseline prompt. Panel~B uses two prompt-based debiasing interventions: a rational investor instruction and an extrapolation warning prepended to the baseline prompt. Error bars denote the 95\% confidence interval.
}
}
\bigskip
\begin{subfigure}[t]{\textwidth}
\caption*{\textbf{Panel A.} Base Model}
\centering
\includegraphics[width=0.8\textwidth]{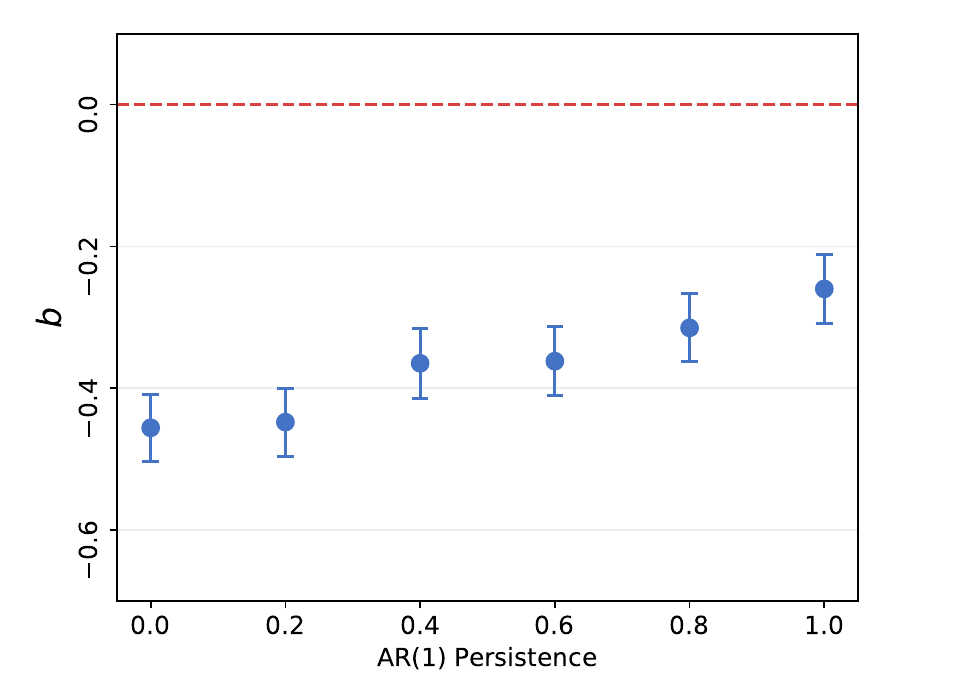}
\end{subfigure}

\vspace{0.5cm}

\begin{subfigure}[t]{\textwidth}
\caption*{\textbf{Panel B.} Prompt-Based Debiasing}
\centering
\includegraphics[width=0.8\textwidth]{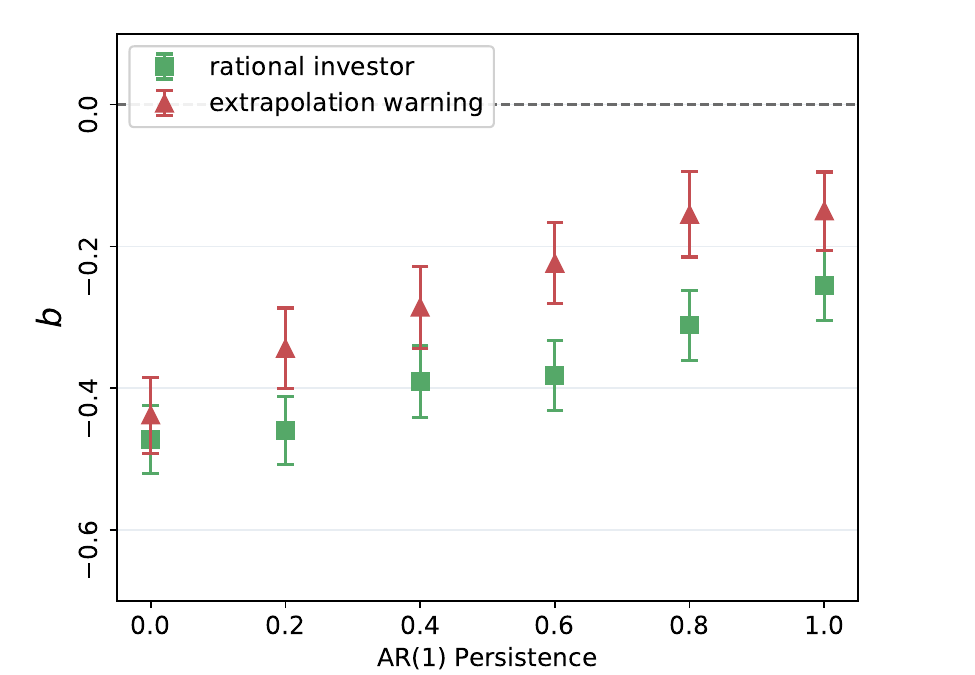}
\end{subfigure}
\end{figure}

\paragraph{Baseline results.} We estimate the error-revision regression in Equation~\eqref{eq:err_rev} separately for each level of~$\rho$ using the LLM generated forecasts.
Table~\ref{tab:descriptive_stats_simulation} reports descriptive statistics for the predicted values. Panel~A of Figure~\ref{fig:qje_table1_replication} and Panel~A of Table~\ref{tab:qje_table1_replication} report the results.
The estimated coefficient~$b$ is negative and statistically significant at the 1\% level for all six persistence conditions, confirming that the language model's forecasts systematically overreact to recent information.
Moreover, the magnitude of overreaction varies with process persistence in the same direction as in the human data: $b$ is most negative at $\rho = 0.0$ ($\hat{b} = -0.456$, $t = -19.05$) and becomes monotonically less negative as persistence increases, reaching $\hat{b} = -0.260$ ($t = -10.37$) at $\rho = 1.0$.
This monotonic pattern is the central empirical finding of \citet{afrouzi2023overreaction}, and the LLM replicates it qualitatively: overreaction is strongest for transitory processes and weakest for random walks.

\paragraph{Prompting alone is insufficient.} \citet{chen2024does} and \citet{bini2025behavioral} show that behavioral biases in LLMs can be moderated through appropriate prompting, but that no single intervention works universally and effective debiasing requires testing multiple prompts. Motivated by this observation, we evaluate two prompt variants before turning to fine-tuning. We first prepend the instruction ``You are a sophisticated rational investor.'' to the baseline prompt and leave the rest of the forecasting task unchanged. We then re-estimate Equation~\eqref{eq:err_rev} on the same test set. The rational-investor series in Panel~B of Figure~\ref{fig:qje_table1_replication} and Panel~B of Table~\ref{tab:qje_table1_replication} report the results. The estimated coefficient on the forecast revision remains negative and significant at the 1\% level for all six persistence conditions, ranging from $\hat{b} = -0.472$ ($t = -19.26$) at $\rho = 0.0$ to $\hat{b} = -0.255$ ($t = -10.23$) at $\rho = 1.0$, closely tracking the baseline pattern in Table~\ref{tab:qje_table1_replication}. Instructing the model to behave as a rational investor therefore does not correct the bias, consistent with the hypothesis that overreaction is encoded in the model's learned parameters rather than in how the forecasting task is framed.

We next consider a more targeted intervention that names the specific bias the model is susceptible to. We prepend the following paragraph to the baseline prompt: ``Extrapolation bias refers to the cognitive tendency to assume recent trends will continue into the future, giving disproportionate weight to the most recent data points while underweighting longer-term patterns, base rates, or the possibility of reversion. Avoid extrapolation bias when creating your response.'' The extrapolation-warning series in Panel~B of Figure~\ref{fig:qje_table1_replication} and Panel~C of Table~\ref{tab:qje_table1_replication} report the results. The warning weakens the magnitude of overreaction relative to the baseline, with $\hat{b}$ falling from $-0.456$ to $-0.438$ at $\rho = 0.0$ and from $-0.260$ to $-0.150$ at $\rho = 1.0$, but the coefficient remains negative and significant at the 1\% level in every persistence condition. Explicitly warning the model against extrapolation therefore reduces the bias but does not eliminate it. Taken together, the two prompt variants confirm the finding of \citet{chen2024does} and \citet{bini2025behavioral} that prompting can lessen but not eliminate behavioral biases in LLMs.

\paragraph{Training dataset and validation dataset.}

To construct a benchmark against which the baseline LLM behavior can be compared, we curate separate training and validation datasets. The training dataset is used to fine-tune the LLM toward the rational expectations forecast, while the validation dataset serves to monitor out-of-sample performance during the fine-tuning process and to guard against overfitting.

The training dataset comprises a large number of round-level observations, constructed from multiple independent AR(1) realizations for each of the six persistence values. The validation dataset is constructed with the same dimensions as the test set. For each observation in both the training and validation datasets, we define learning targets that encode the conditionally optimal forecast.

\paragraph{Fine-tuned LLM corrects the bias.}

The preceding analysis establishes that the pretrained LLM systematically overreacts to recent information, replicating the central finding of \citet{afrouzi2023overreaction} with human subjects. A natural question is whether this bias reflects a fundamental limitation of the model architecture or whether it can be corrected through supervised learning on rational expectations targets. To investigate, we fine-tune Qwen3-32B on the training dataset described above and evaluate the resulting model on the same held-out test set.

\begin{figure}[H]
\centering
\caption{Controlled Experiment: Fine-tuned LLM}
\label{fig:qje_table2_finetuned}
\vspace{0.2cm}
{\fontsize{10}{13}\selectfont
\parbox{\textwidth}{
This figure plots the estimated overreaction coefficient $b$ from the error-revision regression $x_{t+1} - F_{i,t}\, x_{t+1} = a + b\,(F_{i,t}\, x_{t+1} - F_{i,t-1}\, x_{t+1}) + v_{i,t}$ for each persistence parameter $\rho \in \{0, 0.2, 0.4, 0.6, 0.8, 1\}$, using forecasts from the fine-tuned Qwen3-32B. Error bars denote the 95\% confidence interval.
}
}
\bigskip
\includegraphics[width=0.8\textwidth]{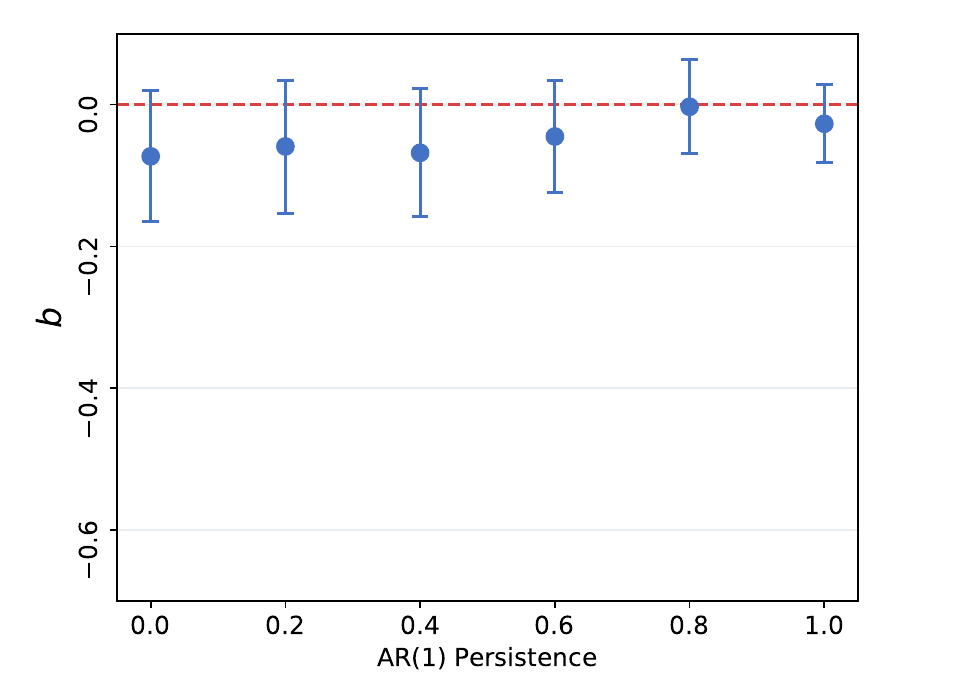}
\end{figure}

Figure~\ref{fig:qje_table2_finetuned} and Table~\ref{tab:qje_table2_finetuned} present the estimates of the error-revision regression in Equation~\eqref{eq:err_rev} using forecasts from the fine-tuned model. The overreaction bias documented in the baseline specification is no longer statistically significant. Across all six persistence conditions, the estimated coefficient on the forecast revision is small in magnitude and statistically indistinguishable from zero at conventional significance levels. Point estimates range from $\hat{b} = -0.073$ ($t = -1.54$) at $\rho = 0.0$ to $\hat{b} = -0.027$ ($t = -0.97$) at $\rho = 1.0$, with none exceeding the 10\% critical value. 
Taken together, these results indicate that the overreaction bias exhibited is a learned regularity that can be corrected through fine-tuning on rational targets.

\subsection{Stock Return Prediction}

\paragraph{Prior literature.} We now turn from forecasting synthetic AR(1) processes to forecasting actual stock returns in the cross-section. In this exercise, we generally follow the approach of~\cite{da2021extrapolative} and~\cite{chen2024does}. \citet{da2021extrapolative} use novel data from the Forcerank platform, a crowdsourcing contest in which participants rank stocks by their expected returns over the coming week, to study how investors form beliefs about short-horizon returns. They document that investors extrapolate from stocks' recent past returns, placing greater weight on more recent realizations. \citet{chen2024does} adopt the same experimental framework, replacing human participants with ChatGPT-4. They find that LLM generated rankings load positively on lagged returns with a pronounced recency effect, mirroring the pattern documented in human forecasts. 

\paragraph{Our setting.} Our analysis departs from both studies in the forecasting target. While \citet{da2021extrapolative} and \citet{chen2024does} focus on the Forcerank contest setting, where participants rank a curated set of stocks, we instead ask the LLM to forecast monthly returns for S\&P 500 constituents. This shift in scope reflects both a practical constraint and a substantive advantage. Because we do not have access to the proprietary Forcerank platform data, we construct our own forecasting exercise using WRDS CRSP for S\&P 500 constituents. The S\&P 500 offers a natural compromise: it captures approximately 80\% of total U.S. equity market capitalization and spans the full range of sectors represented in the broader market, while keeping the prompting budget tractable. We note that our methodology is readily extensible to the full cross-section of listed equities as inference costs decline.

\paragraph{Prompt design and regression specification.} As in the AR(1) exercise, we use Qwen3-32B as the forecasting agent, replacing human participants in the original experimental setting. %
For each stock $i$ in each month $t$, we provide the model with a trailing window of monthly returns and ask it to forecast the stock's return over the following month, $r_{i,t+1}$. As in the AR(1) exercise, we anonymize the prompt by supplying only numerical return data with no firm identifying information and no date information, thereby mitigating potential lookahead bias from the model's pretraining corpus.\footnote{A growing body of work has examined the potential for lookahead bias in
LLM-based predictions~\citep[e.g.,][]{glasserman2023assessing,
sarkar2024lookahead, gao2025testLookaheadBias, lopez2025memorization,
ludwig2025large}. Mitigation strategies include entity-neutering prompting
~\citep[e.g.,][]{
glasserman2023assessing, engelberg2025entity} and training models
under controlled information sets~\citep[e.g.,][]{sarkar2024storieslm,
he2025chronologically, he2025instruction, yan2026datedgpt}. We adopt the
former approach, masking firm names and dates to limit data contamination. Moreover, if residual lookahead bias were present, the LLM
would plausibly generate forecasts that earn positive returns rather than
exhibit the extrapolative patterns we document.}

To quantify the degree of extrapolation, we estimate the following regression:
\begin{equation}\label{eq:extrapolation}
    F_{i,t}\, r_{i,t+1} \;=\; \alpha_i \;+\; \delta_t \;+\; \sum_{s=0}^{11} \beta_s \, r_{i,t-s} \;+\; \varepsilon_{i,t},
\end{equation}
where $F_{i,t}\, r_{i,t+1}$ denotes the LLM's forecast of stock $i$'s return in month $t+1$, formed at the end of month $t$, and $r_{i,t-s}$ is the realized return $s$ months prior. A positive coefficient $\beta_s$ indicates that the model extrapolates from past returns when forming its forecast.

\paragraph{Baseline results.} Table~\ref{tab:descriptive_stats_stock} reports descriptive statistics for the key variables. Table~\ref{tab:stock_return_extrapolation} reports the results. The test set uses a data sample from January 2016 to December 2024. Consistent with the findings of \citet{chen2024does}, the LLM exhibits extrapolation from recent past returns when forming beliefs about future stock returns. Column~(1) estimates Equation~\eqref{eq:extrapolation}, regressing the LLM's forecast on trailing monthly returns with firm and month fixed effects. All coefficients on lagged returns are positive and statistically significant at the 1\% level, confirming that the model systematically extrapolates from past performance. The coefficient on the most recent return $r_{i,t}$ is $0.394$ ($t = 53.92$), and the coefficients generally decline with lag length, falling to $0.111$ at lag one, $0.035$ at lag two, and $0.025$ at lag eleven. The estimated coefficients indicate that the model places excessive weight on recent performance, replicating the central pattern documented in human subjects by \citet{da2021extrapolative}.

\paragraph{Prompting alone is insufficient.} Motivated by the same evidence in \citet{chen2024does} and \citet{bini2025behavioral} that behavioral biases in LLMs can be lessened through appropriate prompting, but that effective debiasing requires testing multiple prompts, we run the analogous exercise in the stock-return setting. We first prepend the instruction ``You are a sophisticated rational investor.'' to the baseline prompt, leave the rest of the forecasting task unchanged, and re-estimate Equation~\eqref{eq:extrapolation} on the same sample of S\&P 500 constituents. Column~(1) of Table~\ref{tab:stock_return_prompting} reports the results. All coefficients on lagged returns remain positive and significant at the 1\% level, and each is slightly larger than its baseline counterpart in Column~(1) of Table~\ref{tab:stock_return_extrapolation}: the coefficient on the most recent return $r_{i,t}$ rises from $0.394$ to $0.430$ ($t = 62.05$), and the coefficient at lag one rises from $0.111$ to $0.138$ ($t = 31.34$). Instructing the model to behave as a rational investor therefore does not correct the extrapolation bias and, if anything, slightly amplifies it. This pattern parallels the controlled-experiment finding above and mirrors the conclusion of \citet{chen2024does}, reinforcing the view that the bias is encoded in the model's pretrained parameters rather than in how the forecasting task is framed.

We next apply the targeted extrapolation warning used in the controlled experiments, prepending to the baseline prompt the paragraph that defines extrapolation bias and instructs the model to avoid it. Column~(4) of Table~\ref{tab:stock_return_prompting} reports the results. The warning substantially weakens the weight the model places on recent returns: the coefficient on $r_{i,t}$ falls from $0.394$ in the baseline to $0.234$ ($t = 29.48$), and the coefficient at lag one falls from $0.111$ to $0.070$ ($t = 13.18$). The coefficients at most intermediate lags remain positive and significant at the 1\% level, and only the most distant lag $r_{i,t-11}$ becomes statistically indistinguishable from zero. Explicitly warning the model against extrapolation therefore reduces the bias but does not eliminate it, paralleling the controlled-experiment finding above. Together, the two prompt variants in this subsection likewise confirm the finding of \citet{chen2024does} and \citet{bini2025behavioral} that prompting can moderate but not eliminate behavioral biases in LLMs.

\paragraph{Training dataset and validation dataset.}

As in the AR(1) exercise, we curate separate training and validation datasets to fine-tune the LLM toward a rational benchmark. We divide the sample into three non-overlapping periods following the convention in~\cite{gu2020empirical}: a training sample spanning January 2001 through December 2011, a validation sample spanning January 2012 through December 2015, and the test sample spanning January 2016 through December 2024 on which the baseline results above are estimated. As before, the validation dataset is used to monitor out-of-sample performance during the fine-tuning process and to guard against overfitting.

\paragraph{Fine-tuned LLM corrects the bias.}

The preceding analysis establishes that the pretrained LLM extrapolates from recent past returns when forming beliefs about future stock returns, replicating the central finding of \citet{da2021extrapolative} and \citet{chen2024does}. We now examine whether fine-tuning on realized return data can correct this bias. We fine-tune Qwen3-32B on the training dataset described above and evaluate the resulting model on the same out-of-sample test period spanning January 2016 through December 2024.

Table~\ref{tab:stock_return_bias_correction} presents the results. Column~(1) re-estimates Equation~\eqref{eq:extrapolation}, regressing the fine-tuned model's forecasts on lagged returns with firm and month fixed effects. The extrapolation bias documented in the baseline specification is reversed. The estimated coefficients on lagged returns are uniformly negative, indicating that the fine-tuned model has internalized the weak mean-reverting tendency in short-horizon stock returns rather than extrapolating from recent performance. The coefficient on the contemporaneous return $r_{i,t}$ is $-0.120$ ($t = -23.21$), and the magnitudes generally decrease with lag length, falling to $-0.005$ ($t = -1.56$) at lag eleven. This pattern is consistent with the model having learned from the training data that stocks with recent positive returns tend to experience subsequent reversals.
Taken together, these results indicate that the extrapolation bias exhibited by the pretrained LLM is not a hard-wired feature of the model architecture. Fine-tuning on realized returns corrects the bias and produces forecasts that reflect the empirical return-generating process. 

\section{Conclusion}

As AI agents built on LLMs move from narrow decision support toward autonomous delegation at scale, the behavioral biases encoded in their underlying models become economically consequential.
We document that LLMs exhibit systematic extrapolation bias in both controlled forecasting environments and the cross-section of stock returns, replicating patterns documented in human subjects.
Prompt-level interventions leave these biases largely intact, consistent with the view that the distortion is embedded in the model's internal representations rather than in the surface-level framing of the input.
By intervening directly at the parameter level through supervised fine-tuning on rational benchmark forecasts, we substantially correct overreaction to recent information in AR(1) forecasting tasks and reverse the extrapolative loading on lagged returns in the cross-section of stock returns, with both corrections holding strictly out-of-sample.
The fine-tuning procedure relies on Low-Rank Adaptation (LoRA), which preserves the model's general language understanding while remaining computationally feasible.
More broadly, our approach offers a low-cost and generalizable methodology for aligning LLM behavior with rational benchmarks across economic settings, a prerequisite for the responsible deployment of AI agents in financial decision-making.

\newpage
\bibliography{ref}

\newpage

\begin{landscape}
  \null
  \vspace{1.5cm}
    \begin{figure}[htbp]
      \centering
      \caption{Pretraining and Instructional Tuning Pipeline}
      \includegraphics[width=\linewidth]{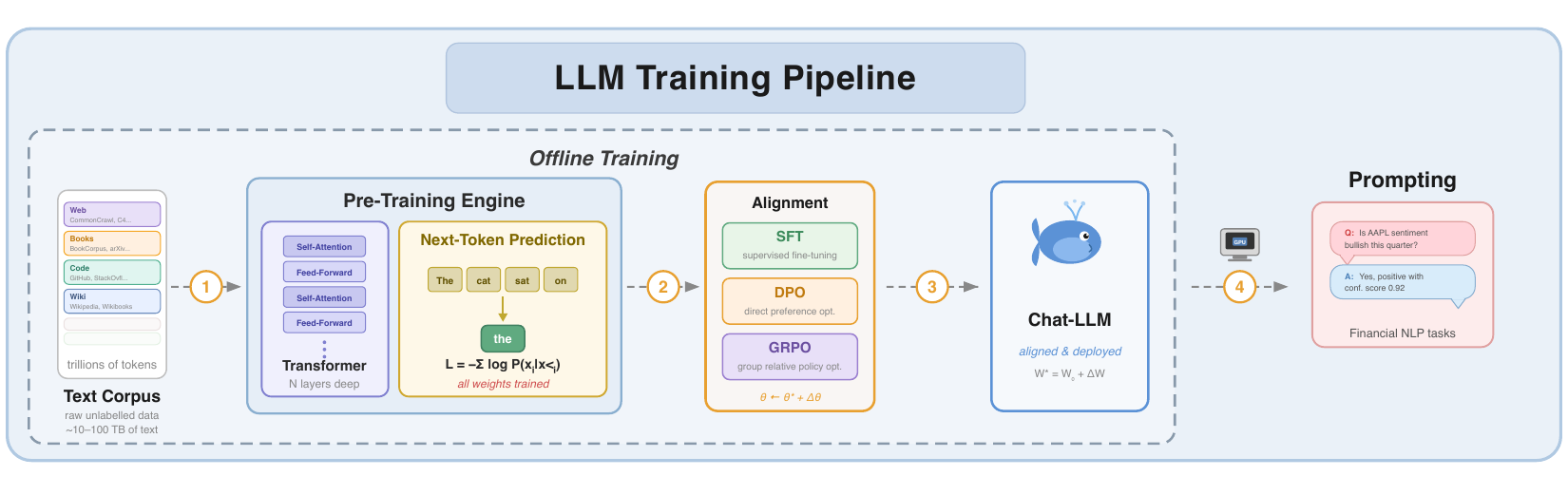}
      \label{fig:pretraining_pipeline}
    \end{figure}
\end{landscape}

\newpage

\begin{landscape}
  \null
  \vspace{1.5cm}
    \begin{figure}[htbp]
      \centering
      \caption{Supervised Fine-Tuning (SFT) Pipeline}
      \includegraphics[width=\linewidth]{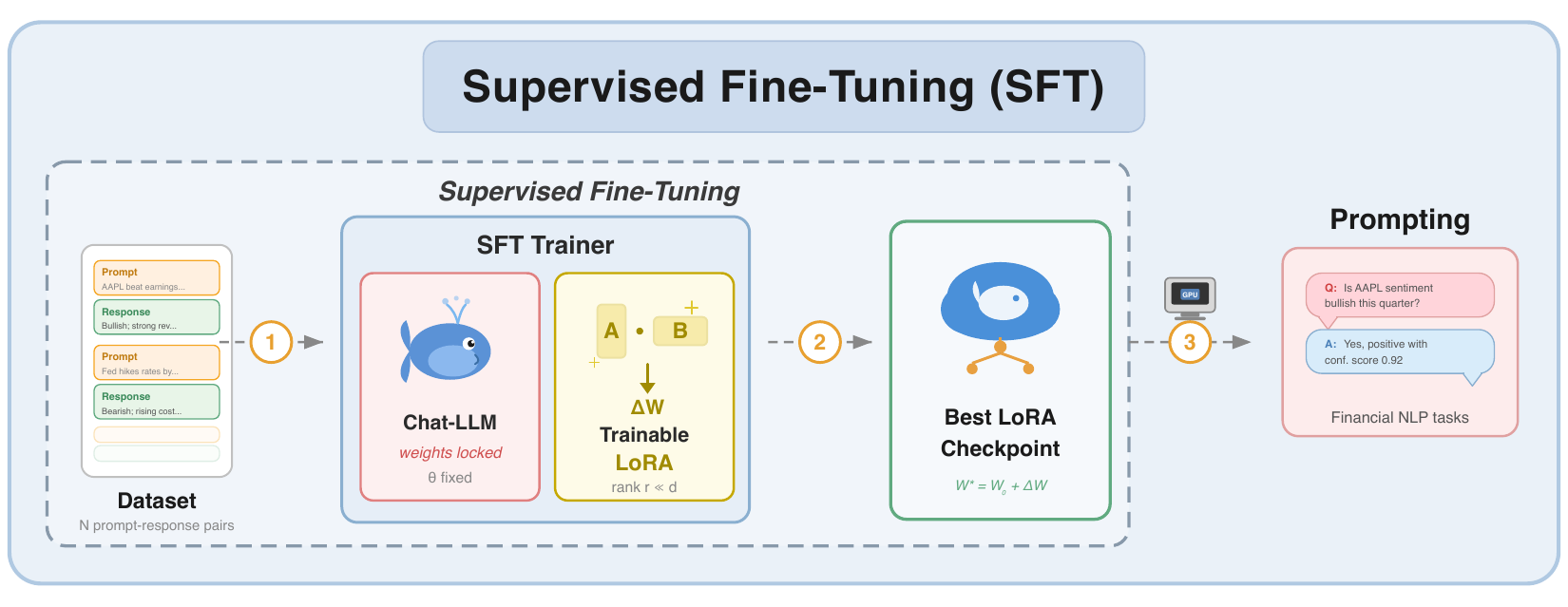}
      \label{fig:sft_pipeline}
    \end{figure}
\end{landscape}

\newpage

\newpage
\begin{landscape}
  \null
  \vspace{1.5cm}
    \begin{figure}[htbp]
      \centering
      \caption{Debiasing Framework}
      \includegraphics[width=\linewidth]{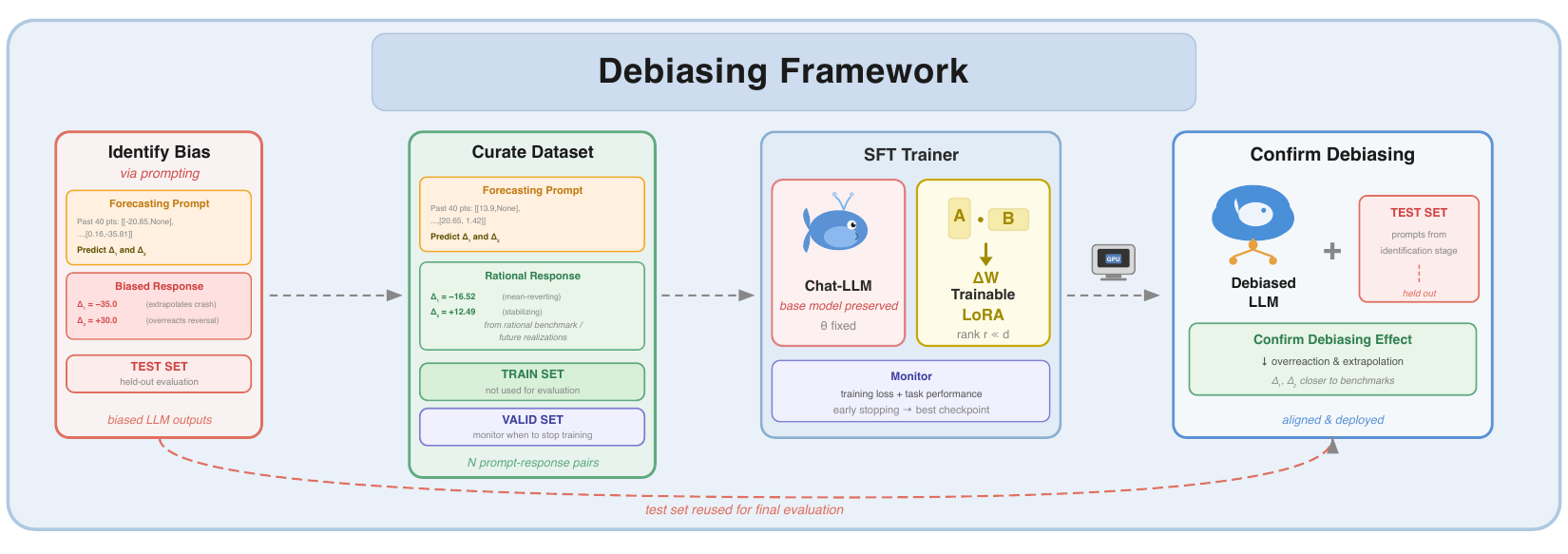}
      \label{fig:debiasing_framework}
    \end{figure}
\end{landscape}

\begin{table}[H]
\caption{Descriptive Statistics: Controlled Experiment}
\label{tab:descriptive_stats_simulation}
  \vspace{0.2cm}
{\fontsize{10}{13}\selectfont
\parbox{\textwidth}{
This table presents descriptive statistics for the predicted values in the AR(1) exercise. $F_{i,t}\, x_{t+1}$ and $F_{i,t-1}\, x_{t+1}$ denote the one-period-ahead and two-period-ahead predictions, respectively. Statistics are reported separately for the base model (Qwen3-32B) and the fine-tuned model across each value of $\rho$. Each cell contains 1,248 observations.
  }
  }
\begin{center}
\resizebox{0.595\textwidth}{!}{%
\begin{tabular}{cl*{5}{c}}
\toprule
$\rho$ & Variable & Mean & SD & P25 & Median & P75 \\
\midrule
\multicolumn{7}{l}{\textit{Panel A: Base Model (Qwen3-32B)}} \\
\addlinespace[3pt]
0.0 & $F_{i,t}\, x_{t+1}$   & $-$0.03 & 18.89 & $-$10.45 & 2.15 & 10.52 \\
    & $F_{i,t-1}\, x_{t+1}$ & 2.99 & 21.22 & $-$10.94 & 2.70 & 17.79 \\
\addlinespace[3pt]
0.2 & $F_{i,t}\, x_{t+1}$   & $-$0.62 & 21.05 & $-$13.07 & 2.37 & 11.92 \\
    & $F_{i,t-1}\, x_{t+1}$ & 2.60 & 21.71 & $-$12.54 & 2.80 & 17.69 \\
\addlinespace[3pt]
0.4 & $F_{i,t}\, x_{t+1}$   & 0.07 & 21.30 & $-$12.95 & 2.15 & 12.93 \\
    & $F_{i,t-1}\, x_{t+1}$ & 2.01 & 23.00 & $-$14.10 & 2.39 & 18.02 \\
\addlinespace[3pt]
0.6 & $F_{i,t}\, x_{t+1}$   & 0.98 & 24.88 & $-$14.87 & 3.03 & 16.38 \\
    & $F_{i,t-1}\, x_{t+1}$ & 1.69 & 25.56 & $-$15.73 & 2.65 & 19.30 \\
\addlinespace[3pt]
0.8 & $F_{i,t}\, x_{t+1}$   & 0.87 & 34.53 & $-$23.45 & 3.51 & 22.67 \\
    & $F_{i,t-1}\, x_{t+1}$ & 1.13 & 35.20 & $-$23.96 & 2.63 & 25.78 \\
\addlinespace[3pt]
1.0 & $F_{i,t}\, x_{t+1}$   & $-$31.63 & 164.32 & $-$132.41 & $-$43.38 & 38.45 \\
    & $F_{i,t-1}\, x_{t+1}$ & $-$33.96 & 164.34 & $-$131.97 & $-$46.24 & 38.74 \\
\addlinespace[6pt]
\midrule
\multicolumn{7}{l}{\textit{Panel B: Fine-tuned Model}} \\
\addlinespace[3pt]
0.0 & $F_{i,t}\, x_{t+1}$   & 0.17 & 3.99 & $-$0.01 & 0.00 & 0.77 \\
    & $F_{i,t-1}\, x_{t+1}$ & 6.02 & 12.08 & $-$1.29 & 10.13 & 14.86 \\
\addlinespace[3pt]
0.2 & $F_{i,t}\, x_{t+1}$   & 0.05 & 6.14 & $-$3.07 & 0.00 & 3.06 \\
    & $F_{i,t-1}\, x_{t+1}$ & 6.19 & 11.79 & $-$1.76 & 7.01 & 14.90 \\
\addlinespace[3pt]
0.4 & $F_{i,t}\, x_{t+1}$   & 0.03 & 9.52 & $-$5.13 & 0.00 & 5.21 \\
    & $F_{i,t-1}\, x_{t+1}$ & 5.97 & 12.18 & $-$2.63 & 4.28 & 15.03 \\
\addlinespace[3pt]
0.6 & $F_{i,t}\, x_{t+1}$   & $-$0.03 & 14.79 & $-$9.66 & 0.09 & 9.17 \\
    & $F_{i,t-1}\, x_{t+1}$ & 4.23 & 15.97 & $-$7.22 & 1.78 & 14.82 \\
\addlinespace[3pt]
0.8 & $F_{i,t}\, x_{t+1}$   & $-$0.10 & 25.85 & $-$17.78 & 0.80 & 16.82 \\
    & $F_{i,t-1}\, x_{t+1}$ & 3.06 & 26.49 & $-$15.04 & 1.19 & 19.80 \\
\addlinespace[3pt]
1.0 & $F_{i,t}\, x_{t+1}$   & $-$32.55 & 157.08 & $-$124.09 & $-$41.54 & 32.07 \\
    & $F_{i,t-1}\, x_{t+1}$ & $-$27.95 & 157.10 & $-$116.35 & $-$38.00 & 34.58 \\
\bottomrule
\end{tabular}%
}
\end{center}
\end{table}

\begin{table}[H]
\centering
\begin{center}
\caption{Controlled Experiment: Base Model and Prompt-Based Debiasing\label{tab:qje_table1_replication}}
\vspace{0.2cm}
{\fontsize{10}{13}\selectfont
\parbox{0.85\textwidth}{
This table reports estimates of the error-revision regression $x_{t+1} - F_{i,t}\, x_{t+1} = a + b\,(F_{i,t}\, x_{t+1} - F_{i,t-1}\, x_{t+1}) + v_{i,t}$ using forecasts from Qwen3-32B. Panel~A uses the baseline prompt. Panel~B prepends the instruction ``You are a sophisticated rational investor.'' to the baseline prompt. Panel~C prepends a paragraph defining extrapolation bias and instructing the model to avoid it. Each column corresponds to a different persistence parameter $\rho$. A negative coefficient $b$ indicates overreaction. $t$-statistics are reported in parentheses. ***, **, and * denote significance at the 1\%, 5\%, and 10\% levels, respectively.
}
}
\\
\vspace{0.25cm}
\resizebox{0.85\textwidth}{!}{%
\begin{tabular}{lcccccc}
\toprule
                & (1)          & (2)          & (3)          & (4)          & (5)          & (6)          \\
                & $\rho=0.0$   & $\rho=0.2$   & $\rho=0.4$   & $\rho=0.6$   & $\rho=0.8$   & $\rho=1.0$   \\
\midrule
\multicolumn{7}{l}{\textbf{Panel A.} Base Model} \\
$F_{i,t}\, x_{t+1} - F_{i,t-1}\, x_{t+1}$ & $-$0.456***  & $-$0.448***  & $-$0.365***  & $-$0.362***  & $-$0.315***  & $-$0.260***  \\
                & ($-$19.05)   & ($-$18.38)   & ($-$14.35)   & ($-$14.52)   & ($-$12.88)   & ($-$10.37)   \\
\addlinespace
$R^2$           & 0.226        & 0.213        & 0.142        & 0.145        & 0.118        & 0.079        \\
$N$             & 1,248        & 1,248        & 1,248        & 1,248        & 1,248        & 1,248        \\
\midrule
\multicolumn{7}{l}{\textbf{Panel B.} Rational Investor Prompt} \\
$F_{i,t}\, x_{t+1} - F_{i,t-1}\, x_{t+1}$ & $-$0.472***  & $-$0.460***  & $-$0.391***  & $-$0.382***  & $-$0.311***  & $-$0.255***  \\
                & ($-$19.26)   & ($-$18.76)   & ($-$15.06)   & ($-$15.29)   & ($-$12.31)   & ($-$10.23)   \\
\addlinespace
$R^2$           & 0.230        & 0.220        & 0.154        & 0.158        & 0.108        & 0.078        \\
$N$             & 1,248        & 1,248        & 1,248        & 1,248        & 1,248        & 1,248        \\
\midrule
\multicolumn{7}{l}{\textbf{Panel C.} Extrapolation Warning Prompt} \\
$F_{i,t}\, x_{t+1} - F_{i,t-1}\, x_{t+1}$ & $-$0.438***  & $-$0.344***  & $-$0.286***  & $-$0.224***  & $-$0.155***  & $-$0.150***  \\
                & ($-$16.00)   & ($-$11.85)   & ($-$9.66)    & ($-$7.62)    & ($-$5.00)    & ($-$5.31)    \\
\addlinespace
$R^2$           & 0.170        & 0.101        & 0.070        & 0.045        & 0.020        & 0.022        \\
$N$             & 1,248        & 1,248        & 1,248        & 1,248        & 1,248        & 1,248        \\
\bottomrule
\end{tabular}%
}
\end{center}
\end{table}

\begin{table}[H]
\centering
\begin{center}
\caption{Controlled Experiment: Fine-tuned LLM\label{tab:qje_table2_finetuned}}
\vspace{0.2cm}
{\fontsize{10}{13}\selectfont
\parbox{0.85\textwidth}{
This table reports estimates of the error-revision regression $x_{t+1} - F_{i,t}\, x_{t+1} = a + b\,(F_{i,t}\, x_{t+1} - F_{i,t-1}\, x_{t+1}) + v_{i,t}$ using forecasts from the fine-tuned Qwen3-32B. Each column corresponds to a different persistence parameter $\rho$. A negative coefficient $b$ indicates overreaction. $t$-statistics are reported in parentheses. ***, **, and * denote significance at the 1\%, 5\%, and 10\% levels, respectively.
}
}
\bigskip
\resizebox{0.85\textwidth}{!}{%
\begin{tabular}{lcccccc}
\toprule
                & (1)          & (2)          & (3)          & (4)          & (5)          & (6)          \\
                & $\rho=0.0$   & $\rho=0.2$   & $\rho=0.4$   & $\rho=0.6$   & $\rho=0.8$   & $\rho=1.0$   \\
\midrule
$F_{i,t}\, x_{t+1} - F_{i,t-1}\, x_{t+1}$ & $-$0.073     & $-$0.059     & $-$0.068     & $-$0.045     & $-$0.003     & $-$0.027     \\
                & ($-$1.54)    & ($-$1.24)    & ($-$1.47)    & ($-$1.12)    & ($-$0.08)    & ($-$0.97)    \\
\addlinespace
$R^2$           & 0.002        & 0.001        & 0.002        & 0.001        & 0.000        & 0.001        \\
$N$             & 1,248        & 1,248        & 1,248        & 1,248        & 1,248        & 1,248        \\
\bottomrule
\end{tabular}%
}
\end{center}
\end{table}

\begin{table}[H]
\caption{Descriptive Statistics: Stock Return Prediction}
\label{tab:descriptive_stats_stock}
  \vspace{0.2cm}
{\fontsize{10}{13}\selectfont
\parbox{\textwidth}{
This table presents descriptive statistics for key variables used in the stock return prediction analysis. Statistics reported include mean, standard deviation (SD), percentiles (P25, Median, P75), and the number of observations (N).
  }
  }

\begin{center}
\resizebox{0.7\textwidth}{!}{%
\begin{tabular}{l*{6}{c}}
\toprule
Variable & Mean & SD & P25 & Median & P75 & N \\
\midrule
$r_{i,t+1}$           & 0.011 & 0.089 & $-$0.038 & 0.011 & 0.060 & 51,724 \\
$F_{i,t}\, r_{i,t+1}$ (Base)           & 0.011 & 0.045 & $-$0.020 & 0.020 & 0.036 & 51,724 \\
$F_{i,t}\, r_{i,t+1}$ (Fine-tuned)     & 0.029 & 0.022 &    0.010 & 0.020 & 0.030 & 51,724 \\
\bottomrule
\end{tabular}%
}
\end{center}
\end{table}

\begin{table}[H]
\centering
\caption{\label{tab:stock_return_extrapolation} Stock Return Extrapolation: Base Model}
\vspace{0.2cm}
{\fontsize{10}{13}\selectfont
\parbox{\textwidth}{
This table examines extrapolation in the base model (Qwen3-32B) stock return forecasts. The sample covers S\&P 500 constituents from January 2016 through December 2024. Column~(1) estimates $F_{i,t}\, r_{i,t+1} = \alpha_i + \delta_t + \sum_{s=0}^{11} \beta_s \, r_{i,t-s} + \varepsilon_{i,t}$, regressing the LLM's forecast on lagged returns with firm and month fixed effects. For brevity, we report coefficients for $s \in \{0, 1, 2, 3, 5, 7, 9, 11\}$. Column~(2) regresses realized returns on lagged returns. Column~(3) regresses realized returns on the LLM's forecast with firm and month fixed effects. Standard errors are double-clustered by stock and month. $t$-statistics are reported in parentheses. ***, **, and * denote significance at the 1\%, 5\%, and 10\% levels, respectively.
}
}
\bigskip
\begin{center}
{\fontsize{11}{13}\selectfont
\begin{tabular}{lccc}
\toprule
                                      & (1) & (2) & (3) \\
                                      & \scalebox{1.1}{$F_{i,t}\, r_{i,t+1}$} & \scalebox{1.2}{$r_{i,t+1}$} & \scalebox{1.2}{$r_{i,t+1}$} \\
\midrule
\scalebox{1.1}{$F_{i,t}\, r_{i,t+1}$} &  &  & $-$0.074$^{**}$ \\
                                      &  &  & ($-$2.05) \\
\scalebox{1.2}{$r_{i,t}$}            & 0.394$^{***}$ & $-$0.048$^{*}$ & \\
                                      & (53.92) & ($-$1.85) & \\
\scalebox{1.2}{$r_{i,t-1}$}          & 0.111$^{***}$ & $-$0.020 & \\
                                      & (27.31) & ($-$1.14) & \\
\scalebox{1.2}{$r_{i,t-2}$}          & 0.035$^{***}$ & $-$0.026 & \\
                                      & (12.98) & ($-$1.16) & \\
\scalebox{1.2}{$r_{i,t-3}$}          & 0.038$^{***}$ & $-$0.017 & \\
                                      & (13.16) & ($-$0.94) & \\
\scalebox{1.2}{$r_{i,t-5}$}          & 0.027$^{***}$ & $-$0.009 & \\
                                      & (11.73) & ($-$0.50) & \\
\scalebox{1.2}{$r_{i,t-7}$}          & 0.018$^{***}$ & $-$0.038 & \\
                                      & (6.88) & ($-$1.50) & \\
\scalebox{1.2}{$r_{i,t-9}$}          & 0.024$^{***}$ & $-$0.008 & \\
                                      & (7.75) & ($-$0.39) & \\
\scalebox{1.2}{$r_{i,t-11}$}         & 0.025$^{***}$ & $-$0.019 & \\
                                      & (7.17) & ($-$1.28) & \\
\midrule
Controls                                & Yes & Yes & No \\
Firm FE                                & Yes & Yes & Yes \\
Month FE                               & Yes & Yes & Yes \\
Estimation                             & Panel & Panel & Panel \\
Within $R^2$                           & 0.632 & 0.007 & 0.001 \\
$N$                                    & 51,724 & 51,724 & 51,724 \\
\bottomrule
\end{tabular}
}
\end{center}
\end{table}

\begin{table}[H]
\centering
\caption{\label{tab:stock_return_prompting} Stock Return Extrapolation: Prompt-Based Debiasing}
\vspace{0.2cm}
{\fontsize{10}{13}\selectfont
\parbox{\textwidth}{
This table examines extrapolation in the stock return forecasts of Qwen3-32B under two prompt-based debiasing interventions. Columns~(1)--(3) prepend the instruction ``You are a sophisticated rational investor.'' to the baseline prompt. Columns~(4)--(6) prepend a paragraph defining extrapolation bias and instructing the model to avoid it. The sample covers S\&P 500 constituents from January 2016 through December 2024. Columns~(1) and~(4) estimate $F_{i,t}\, r_{i,t+1} = \alpha_i + \delta_t + \sum_{s=0}^{11} \beta_s \, r_{i,t-s} + \varepsilon_{i,t}$, regressing the LLM's forecast on lagged returns with firm and month fixed effects. For brevity, we report coefficients for $s \in \{0, 1, 2, 3, 5, 7, 9, 11\}$. Columns~(2) and~(5) regress realized returns on the LLM's forecast with firm and month fixed effects. Columns~(3) and~(6) repeat columns~(2) and~(5) without firm fixed effects, clustering standard errors by month only. Standard errors in columns~(1), (2), (4), and~(5) are double-clustered by stock and month. $t$-statistics are reported in parentheses. ***, **, and * denote significance at the 1\%, 5\%, and 10\% levels, respectively.
}
}
\vspace{0.5cm}
\begin{center}
{\fontsize{9}{11}\selectfont
\setlength{\tabcolsep}{4pt}
\begin{tabular}{lcccccc}
\toprule
                   & \multicolumn{3}{c}{Rational Investor} & \multicolumn{3}{c}{Extrapolation Warning} \\
\cmidrule(lr){2-4} \cmidrule(lr){5-7}
                   & (1) & (2) & (3) & (4) & (5) & (6) \\
                   & $F_{i,t}\, r_{i,t+1}$ & $r_{i,t+1}$ & $r_{i,t+1}$ & $F_{i,t}\, r_{i,t+1}$ & $r_{i,t+1}$ & $r_{i,t+1}$ \\
\midrule
$F_{i,t}\, r_{i,t+1}$ &  & $-$0.068$^{**}$ & $-$0.029 &  & $-$0.098$^{**}$ & $-$0.039 \\
                   &  & ($-$2.02) & ($-$0.86) &  & ($-$2.14) & ($-$0.78) \\
$r_{i,t}$          & 0.430$^{***}$ & & & 0.234$^{***}$ & & \\
                   & (62.05) & & & (29.48) & & \\
$r_{i,t-1}$        & 0.138$^{***}$ & & & 0.070$^{***}$ & & \\
                   & (31.34) & & & (13.18) & & \\
$r_{i,t-2}$        & 0.042$^{***}$ & & & 0.022$^{***}$ & & \\
                   & (14.41) & & & (6.64) & & \\
$r_{i,t-3}$        & 0.047$^{***}$ & & & 0.031$^{***}$ & & \\
                   & (14.54) & & & (11.17) & & \\
$r_{i,t-5}$        & 0.028$^{***}$ & & & 0.019$^{***}$ & & \\
                   & (9.29) & & & (8.25) & & \\
$r_{i,t-7}$        & 0.022$^{***}$ & & & 0.012$^{***}$ & & \\
                   & (9.14) & & & (5.35) & & \\
$r_{i,t-9}$        & 0.019$^{***}$ & & & 0.019$^{***}$ & & \\
                   & (6.42) & & & (8.20) & & \\
$r_{i,t-11}$       & 0.029$^{***}$ & & & $-$0.001 & & \\
                   & (8.38) & & & ($-$0.59) & & \\
\midrule
Controls           & Yes & No & No & Yes & No & No \\
Firm FE            & Yes & Yes & No & Yes & Yes & No \\
Month FE           & Yes & Yes & Yes & Yes & Yes & Yes \\
Estimation         & Panel & Panel & Panel & Panel & Panel & Panel \\
Within $R^2$       & 0.617 & 0.002 & 0.0003 & 0.496 & 0.001 & 0.0002 \\
$N$                & 51,724 & 51,724 & 51,724 & 51,724 & 51,724 & 51,724 \\
\bottomrule
\end{tabular}
}
\end{center}
\end{table}

\begin{table}[H]
\centering
\caption{\label{tab:stock_return_bias_correction} Stock Return Extrapolation: Fine-tuned LLM}
\vspace{0.2cm}
{\fontsize{10}{13}\selectfont
\parbox{\textwidth}{
This table examines extrapolation in the fine-tuned Qwen3-32B's stock return forecasts. The sample covers S\&P 500 constituents from January 2016 through December 2024 . The model is fine-tuned on realized returns from January 2001 through December 2011, with January 2012 through December 2015 used for validation. Column~(1) estimates $F_{i,t}\, r_{i,t+1} = \alpha_i + \delta_t + \sum_{s=0}^{11} \beta_s \, r_{i,t-s} + \varepsilon_{i,t}$, regressing the fine-tuned LLM's forecast on lagged returns with firm and month fixed effects. For brevity, we report coefficients for $s \in \{0, 1, 2, 3, 5, 7, 9, 11\}$. Column~(2) regresses realized returns on the fine-tuned LLM's forecast with firm and month fixed effects. Standard errors are double-clustered by stock and month. $t$-statistics are reported in parentheses. ***, **, and * denote significance at the 1\%, 5\%, and 10\% levels, respectively.
}
}
\bigskip
\begin{center}
{\fontsize{11}{13}\selectfont
\begin{tabular}{lcc}
\toprule
                                      & (1) & (2) \\
                                      & \scalebox{1.1}{$F_{i,t}\, r_{i,t+1}$} & \scalebox{1.2}{$r_{i,t+1}$} \\
\midrule
\scalebox{1.1}{$F_{i,t}\, r_{i,t+1}$} &  & 0.136$^{**}$ \\
                                      &  & (2.26) \\
\scalebox{1.2}{$r_{i,t}$}            & $-$0.120$^{***}$ & \\
                                      & ($-$23.21) & \\
\scalebox{1.2}{$r_{i,t-1}$}          & $-$0.062$^{***}$ & \\
                                      & ($-$20.37) & \\
\scalebox{1.2}{$r_{i,t-2}$}          & $-$0.037$^{***}$ & \\
                                      & ($-$16.18) & \\
\scalebox{1.2}{$r_{i,t-3}$}          & $-$0.022$^{***}$ & \\
                                      & ($-$7.40) & \\
\scalebox{1.2}{$r_{i,t-5}$}          & $-$0.011$^{***}$ & \\
                                      & ($-$5.50) & \\
\scalebox{1.2}{$r_{i,t-7}$}          & $-$0.012$^{***}$ & \\
                                      & ($-$6.41) & \\
\scalebox{1.2}{$r_{i,t-9}$}          & $-$0.002 & \\
                                      & ($-$0.99) & \\
\scalebox{1.2}{$r_{i,t-11}$}         & $-$0.005 & \\
                                      & ($-$1.56) & \\
\midrule
Controls                                & Yes & No \\
Firm FE                                & Yes & Yes \\
Month FE                               & Yes & Yes \\
Estimation                             & Panel & Panel \\
Within $R^2$                           & 0.301 & 0.001 \\
$N$                                    & 51,724 & 51,724 \\
\bottomrule
\end{tabular}
}
\end{center}
\end{table}

\appendix
\section{Online Appendix}
\renewcommand{\thetable}{A.\arabic{table}} %
\setcounter{table}{0}                      %
\begin{table}[ht]
\centering
\caption{Top Dense Text Generation Models on Hugging Face as of March 2026}
\label{tab:text-gen-models}
\resizebox{\columnwidth}{!}{%
\begin{tabular}{lrrr}
\hline
\textbf{Model} & \textbf{Total Params} & \textbf{Active Params} & \textbf{Downloads} \\
\hline
\href{https://huggingface.co/Qwen/Qwen2.5-7B-Instruct}{\textcolor{black}{Qwen/Qwen2.5-7B-Instruct}}        & 8B  & 8B  & 20.9M \\
\href{https://huggingface.co/Qwen/Qwen3-8B}{\textcolor{black}{Qwen/Qwen3-8B}}                    & 8B  & 8B  & 8.97M \\
\href{https://huggingface.co/meta-llama/Llama-3.1-8B-Instruct}{\textcolor{black}{meta-llama/Llama-3.1-8B-Instruct}} & 8B  & 8B  & 7.62M \\
\href{https://huggingface.co/Qwen/Qwen3-32B}{\textcolor{black}{Qwen/Qwen3-32B}}                   & 32B & 32B & 4.81M \\
\href{https://huggingface.co/dphn/dolphin-2.9.1-yi-1.5-34b}{\textcolor{black}{dphn/dolphin-2.9.1-yi-1.5-34b}}    & 34B & 34B & 4.61M \\
\hline
\end{tabular}%
}
\end{table}

\clearpage
\newpage
\end{document}